\documentclass[preprint]{aastex}

\begin{document}

\title{Deep HST Imaging of Sextans A I. 
The Spatially Resolved Recent Star Formation History}

\author{Robbie C. Dohm-Palmer}
\affil{Astronomy Department, University of Michigan, Ann Arbor, MI 
48109}
\email{rdpalmer@astro.lsa.umich.edu}

\author{Evan D. Skillman}
\affil{Astronomy Department, University of Minnesota, Minneapolis, MN 
55455}
\email{skillman@astro.umn.edu}

\author{Mario Mateo}
\affil{Astronomy Department, University of Michigan, Ann Arbor, MI 
48109}
\email{mateo@astro.lsa.umich.edu}

\author{Abi Saha and Andrew Dolphin}
\affil{National Optical Astronomy Observatories, 950 North Cherry 
Avenue, P.O. Box 26732, Tucson, AZ 85726}
\email{saha@noao.edu, dolphin@noao.edu}

\author{Eline Tolstoy}
\affil{U.K. Gemini Support Group, Oxford University, Oxford OX1 3Rh, 
England}
\email{etolstoy@astro.ox.ac.uk}

\author{Jay S. Gallagher}
\affil{Department of Astronomy, University of Wisconsin, Madison, WI 
53706-1582}
\email{jsg@astro.wisc.edu}

\and

\author{Andrew A. Cole}
\affil{Astronomy Dept., 532 LGRT,
University of Massachusetts, Amherst, MA 01003}
\email{cole@condor.astro.umass.edu}

\begin{abstract}

We have measured stellar photometry from deep Cycle~7 Hubble Space
Telescope/WFPC2 imaging of the dwarf irregular galaxy Sextans~A. The
imaging was taken in three filters: F555W ($V$; 8 orbits), F814W ($I$;
16 orbits), and F656N (H$\alpha$; 1 orbit). Combining these data with
Cycle~5 WFPC2 observations provides nearly complete coverage of the
optically visible portion of the galaxy. The Cycle~7 observations are
nearly 2 magnitudes more sensitive than the Cycle~5 observations,
which provides unambiguous separation of the faint blue helium burning
stars (BHeB stars) from contaminant populations.  The depth of the
photometry allows us to compare recent star formation histories
recovered from both the main sequence (MS) stars and the BHeB stars
for the last 300 Myr. The excellent agreement between these
independent star formation rate (SFR) calculations is a resounding
confirmation for the legitimacy of using the BHeB stars to calculate
the recent SFR.  Using the BHeB stars we have calculated the global
star formation history over the past 700 Myr. The history calculated
from the Cycle~7 data is remarkably identical to that calculated from
the Cycle~5 data, implying that both halves of the galaxy formed stars
in concert. We have also calculated the spatially resolved star
formation history, combining the fields from the Cycle~5 and Cycle~7
data. The star forming regions are found in 3 major zones of the
galaxy. One of these zones is extremely young, consisting of only a
single star forming region which is less than 20 Myr old. Two of these
zones are associated with high column density neutral gas, while the
third, and oldest, is not. Our interpretation of this pattern of star
formation is that it is an orderly stochastic process. Star formation
begins on the edge of a gas structure, and progressively eats away at
the cloud, breaking it up and inducing further star formation. A more
quantitative analysis of the star formation process must await a
larger sample of galaxies with spatially resolved star formation
histories to allow correlation studies with the physical properties of
the galaxy.

\end{abstract}


\section{Introduction}

Dwarf galaxies are critical components in the evolution of the
universe. High-redshift observations, and modern simulations of
structure evolution within the framework of cold dark matter
\citep{pea99, ste99}, suggest that large galaxies formed
hierarchically through the progressive merger of smaller pre-galactic
structures. Dwarf galaxies in the nearby universe are obvious
candidates for the remains of these smaller objects \citep{tol99}.
One significant problem with this interpretation is that models
predict there should be far more dwarfs in the nearby universe than we
observe \citep{kkvp99, mgglqst99}.  Furthermore, the merger process
clearly continues today, as evidenced by the Sgr dwarf \citep{iba94},
making the evolutionary role of nearby dwarfs ambiguous.

There are several puzzling aspects of nearby dwarfs that prevent us 
from clearly identifying their evolutionary role.  Do nearby dwarfs 
represent the initial, intermediate, or final stages of hierarchical 
formation?  Why did early-type dwarfs use up or lose all their gas, 
while late-types retain gas and continue to form stars?  What is the 
relation between dI galaxies with low to moderate star formation 
rates, and more ``bursty'' galaxies with currently high star formation 
rates? Answers to these questions may depend critically on the merger 
history of the universe. 

Resolved stellar populations in nearby galaxies can address many of
these questions by directly determining the star formation history for
these galaxies.  However, in order to separate potential merger events
from simple star formation events we must independently understand the
star formation process.  The physical motives for star formation in
dwarf galaxies are very poorly understood.  Unlike spiral galaxies
\citep{r69, kv87}, dwarfs have no spiral density waves to mediate the
formation of molecular clouds.  This lack of global dynamic structure
suggests that stochastic processes will play a significant, if not
dominant, role in dwarfs.

If the star formation is stochastically induced, one might expect the
history of a dwarf to be volatile: filled with large bursts of
activity and lulls of inactivity.  Certainly this characterizes the
class of blue compact dwarf galaxies \citep{ss72, ssb73}.  This type
of behavior has also been demonstrated in several dwarf spheroidal
galaxies whose color-magnitude diagrams (CMD's) clearly show several
major episodes of star formation with little activity in between
\citep{sme94, mig96, m98}. However, the evolutionary role of these
objects is ambiguous. Did these objects undergo this quasi-periodic
behavior as a single entity, or as separate entities that have
subsequently merged?

One might assume that dI galaxies undergo a similar history, and are
the same type of galaxy as the more ``bursty'' objects, caught in a
different phase of evolution.  However, there is little evidence for
pronounced starburst cycles among gas-rich field dwarfs in the present
epoch \citep{van01, gal95}.  The lack of bursts in these isolated objects may
be due to a lack of merger or interaction in their past.  If so, these
isolated dI galaxies may represent the most basic building block of
larger objects.  This is consistent with the recent observation that
isolated objects closely match a closed-box chemical evolution model
\citep{ks01}.

In order to address these evolution processes we continue our study of
star formation in the dI galaxy Sextans~A. Our initial studies
\citep{doh97a, doh97b} were based on a total of 2 orbits of WFPC2
images in the F555W and F814W filters.  This field covered
approximately half the visible portion of the galaxy.  The current
study adds Cycle~7 WFPC2 imaging in these same two filters totaling 24
orbits.  This deeper field was approximately positioned to include the
opposing half of the visible galaxy.  In this paper we explore the
spatially resolved star formation history of Sextans~A.  The next
paper in this series will explore the intermediate to old ($\ge$1 Gyr)
star formation history using the red-giant branch and red clump
populations.

Our techniques for calculating the recent star formation history,
discussed in detail in \citet{doh97b}, utilize the Blue He-Burning
(BHeB) supergiants to calculate both the rate and location of star
formation over the past 1 Gyr.  This technique follows in the
pioneering footsteps of the studies of the Large Magellanic Clouds by
\citet{pg73} and \citet{i84}.  By combining these new Cycle~7 data with
the previous Cycle~5 data, we obtain a nearly complete view of the
visible galaxy.  This allows us to make broader comparisons between
local star formation properties and global properties.  One appeal of
resolved recent star formation histories is that it allows a detailed
insight into the star formation process, allowing an investigation of
the patterns of star formation and a comparison between the
distributions of recently formed stars and the gas from which they
formed \citep{ik84, doh98}.  Studies of many galaxies in this
way will allow us to characterize properties such as typical duration
of star forming episodes and typical sizes of star forming regions.
They will also help us to better understand whether the star formation
process is predominantly stochastic, or whether ``triggers'' are
required.

More complete observational histories of Sextans~A can be found in
\citet{doh97a} and \citet{vdb00}.  Since the study by \citet{doh97a},
a detailed observational study has been conducted by \citet{van98}.
\citet{van98} measured $UBVRI$ CCD stellar photometry using the KPNO
2.1m telescope and compared these to HI imaging and narrow-band
H$\alpha$ imaging. They make extensive spatial comparisons of
different age populations to conclude that star formation has been
propagating outward from the center of the galaxy over the last 100
Myr.  This is imminently relevant to our study, and we make detailed
comparisons between their conclusions and ours in \S 5.3.

Arguably the most important parameter in calculating the star
formation history from stellar populations is the distance
\citep{tol98}.  We adopt the distance used in \citet{doh97a} of 1.44
Mpc ($m-M_0 = 25.8 \pm 0.1$).  This is a compromise between the
cepheid determination and the RGB tip determination
\citep{pcp94,smf96}.  This is also the same distance adopted by
\citet{van98}.

In the following section we describe the new data set and image
construction. This is followed by a discussion of the photometric
measurement. The moderate stellar crowding in these deep images
requires considerable care in modeling the PSF. We then compare the
photometry with the Cycle~5 photometry for the small overlap
region. Next, we describe the global star formation history
calculation. We compare the results of using the main sequence to
using the BHeB stars. We then use the BHeB stars to calculate the
spatially resolved star formation history.  The results of the
spatially resolved history are compared with the conclusions of
\citet{van98}.  A comparison is made with the neutral gas (HI)
distribution.  Finally, we discuss the implications for the star
formation process and the evolutionary role of dI galaxies.

\section{Observations and Image Construction}
\label{obsSec}

The images were taken with the Hubble Space Telescope(HST)/WFPC2
instrument between the dates of 5 April 1999 and 8 April 1999.  Images
in the F555W filter were obtained during 8 orbits, F814W during 16
orbits, and F656N during 1 orbit.  Each orbit consists of two separate
images, each 1200 seconds long.  Thus, the total integration times
were 19200 seconds in F555W, 38400 seconds in F814W, and 2400 seconds
in F656N (Table \ref{tabobs}).  The images were taken in four
different sub-pixel dithering positions.  The orbits were divided
evenly among the four positions. The images were calibrated in the
standard HST pipeline reduction.

The first step is to combine the exposures into a single image for
each filter. The images were shifted to align to the nearest integer
pixel. To determine the shifts for each image, we used the DoPHOT
program (described in more detail below) to measure the position of
several thousand stars in each image. The stars from each image were
matched and the coordinate transformation used to determine the
nearest integer shift.

The images were then coadded, applying an anti-coincidence algorithm
to detect and remove cosmic ray detections between pairs of
images. The algorithm was optimized for the WFPC2 PSF and is described
in \citet{sah96}. All the images were combined in pairs, starting with
the pairs taken in the same orbit, until a single image was obtained.
At each combination cosmic ray detections were removed to only the
4-sigma level. While this potentially leaves some cosmic ray
detections, this procedure is performed many times with successively
less and less noisy images.  Hence, the final images are quite clean
of cosmic ray detections.

The final images are shown in Fig.\ \ref{figwhole}.  The HST images 
are overlaid on a ground-based image taken on the 0.9m at CTIO 
\citep{hun97}.  We have also included the previous HST image of 
Sextans~A, described in \citet{doh97a}.  We recreated the image for 
the old position to include only the F555W and F814W filters, so that 
it would match the colors of the new position. Where the Cycle~5 and
Cycle~7 images overlap, the Cycle~7 image is shown. 

The F656N image (H$\alpha$) at the new position has been added in red.  
Note that the F656N image is much noisier than the other images (as
expected for a narrow band image).  The primary feature in the H$\alpha$ 
is a large semi-circular shell.  This was first seen by \citet{hun92}, 
and later divided into 9 HII regions by \citet{hod94}.

\section{Stellar Photometry}

\subsection{Photometric Measurements}

Stellar photometry was measured from the final F555W and F814W images,
following the procedure outlined by \citet{doh97a}.  The DoPHOT
program \citep{sch93} was used, with modifications for the
under-sampled WFPC2 PSF, as described in \citet{sah96}.  With the depth
achieved in these images, there is a moderate amount of stellar
crowding.  DoPHOT attempts to overcome crowding difficulties by
subtracting an analytic function from the image at the location of
each identified star, thus isolating stars for photometric
measurement.  

In an uncrowded field, it is not important to precisely match the
fitting function to the PSF. However, this is not the case in a
crowded field.  Residuals of the subtraction in neighboring stars can
interact to mimic faint, noisy stars. This can change the photometric
measurement by inadvertently removing flux from the image.  This
difficulty is further enhanced by the fact that the PSF in the WFPC2
images is under-sampled.  Because the PSF covers few pixels, it is
statistically easier for the interacting residuals to conspire to
mimic faint stars. To avoid this problem as much as possible, we
exerted great effort to match the analytic function to the PSF.

The shape of the fitting function in DoPHOT is controlled by 3
parameters: $\beta_4$, $\beta_6$, and $\beta_8$.  These parameters
were determined following the method described in \citet{doh97a}.  In
this method, the parameters are allowed to vary.  For each parameter
set, the function is fit to a group of bright stars, and subtracted
from the image.  The parameter set that leaves the smallest residual
is chosen as the best.  We found that the best overall results were
achieved by using the same set of $\beta$ values for all WF chips and
a different set for the PC, while letting the FWHM vary from chip to
chip.  The values used for this data set are given in Table
\ref{tabpsf}.

Despite the effort, the fitting function does not match the PSF
perfectly, and there is still a need for an aperture correction. The
aperture corrections were measured with an aperture of 0.5\arcsec\
around the brightest stars with internal DoPHOT errors less than 0.04.
A polynomial, quadratic in position, was fit to these values.  Table
\ref{tabaper} lists the aperture correction fits for each chip and
filter.  Also listed are the number of stars used in the fit, and the
RMS residual of the fit.

The first thing to notice is that the quadratic fits were only
slightly better than fitting a constant value. This is reflected in
the small coefficients.  Second, the correction is quite small.  To
demonstrate this, the aperture corrections for the center of each chip
are also listed in Table \ref{tabaper}.  The central value is less
than the RMS residual of the fit in all but one case.  This indicates
we likely could not improve the choice of $\beta$ parameters without
letting them vary spatially across each chip.

\subsection{Photometric Results}

Once the stellar photometry was obtained from the final F555W and
F814W images, the two lists were matched to within a radius of 0.5
pixels.  Prior to calibration, the photometry was corrected for charge
transfer efficiency (CTE) using the equations of \citet{whi99}, which
have been verified by the study of \citet{d00}.  We used the
transformation to $V$ and $I$ in \citet{hol95}.

We adopted the Galactic reddening value of $E(B-V)=0.043$ and the
extinction coefficients given in \citet{sch98}: $A_V = 3.315 E(B-V)$
and $A_I = 1.940 E(B-V)$.  This gave $A_V = 0.143$ and $A_I = 0.083$.
These extinction values are slightly higher than was used in
\citet{doh97a}, however the difference in $E(V-I)$ is only 0.01.  The
low values of extinction are consistent with the high Galactic
latitude of Sextans A ($+$40$^o$) and the low metal abundance of its
ISM \citep{skh89}.  This low overall extinction is probably critical
in the clear separation of the bright MS and BHeB stars.  The low
extinction puts a strong upper limit on the degree of differential
reddening that can broaden the two bright blue star features.  The
importance of this effect has been clearly demonstrated for the Large
Magellanic Cloud \citep{hzt97, z99}.

The internal DoPHOT measurement errors are shown in Fig.\ 
\ref{figerr}.  The $V$ data are slightly deeper than the $I$ data.  
The present data are deeper than the Cycle~5 data by 1.5 mag and 1.8 
mag in $V$ and $I$ respectively.  Most of the points follow the 
standard pattern of broadening and increasing toward fainter 
magnitude.  The scatter of points above this curve comprise about 4\% 
of the data.  This is a slightly higher fraction than was observed in 
the Cycle~5 data, which we attribute to the effects of crowding.

We performed artificial star tests, following the procedure detailed
in \citet{doh97a}.  The PSF was created by averaging between 10 and 20
stars for each of the WF chips, and 4 stars for each of the PC chips.
While this is a small number of stars for the PC, we preferred to
accept only bright, completely isolated stars.  The high stellar
density limited the availability of such stars.  The broadening factor
due to the under-sampling of the PSF (see \citet{doh97a}) was measured
to be $\sim 1.1$, and varied from chip to chip by 4\%.

Artificial stars were added to 100 copies of the final images in both
filters.  Each copy contained artificial stars numbering 2\% of the
detected real stars.  The artificial stars were distributed with a
power law of slope 0.35 in $V$, matching the real data, and a uniform
color distribution with $-1 < V-I < 2$. DoPHOT was run on each frame
identically to the original images. The data were then matched and
calibrated identically, including the CTE correction. Finally, the
star lists were matched with the lists of artificial stars.

The results of the artificial star tests are shown in Fig.\
\ref{figcomp} as plots of incompleteness versus magnitude.  
Errors in the histogram are from Poisson statistics
based on the number of input artificial stars.  As described in
\citet{doh97a} the faint end of the completeness curves needed to be
corrected for Malmquist bias.  Also, the bright end was fairly noisy,
producing undesirable structure.  To avoid introducing this structure
to the later calculations we have boxcar smoothed the curves brighter
than 25 magnitudes.

Notice that for magnitudes between 22 and 26 nearly 20\% of the
artificial stars are missing.  This is an effect of the high stellar
density.  These stars have been blended with existing stars, and thus
were removed from the detection list.  Despite the crowding, the
photometric fidelity of the detected, non-blended, artificial stars
was excellent. The recovered magnitudes agreed with the input
magnitudes consistent with the random noise.

\subsection{Comparison to Cycle 5 Photometry}

There is a small region of overlap between the Cycle~5 and Cycle~7
data sets.  We have compared the photometry for stars in this region
as a check on the photometric consistency.  Approximately 8200 stars
were found in common in this region.  There is a small zero-point
offset between the two data sets ($\sim 0.03$ mag).  In both cases the
offset is such that the Cycle~7 data are brighter. Since the offset is
nearly the same in both $V$ and $I$, the color index $V-I$ is
practically identical for the two data sets.  We believe this small
offset is due to the CTE correction, which has a large time
dependence. \citet{whi99} suggest the correction is only good to $\sim
5$\%. In this case the agreement is good to $\sim 3$\%, which is
probably as good as can be expected.

The CMD's for Cycle~7 and Cycle~5 are shown in Fig.\
\ref{figcmd}. Also plotted in each are curves indicating the main
sequence (MS), the blue edge of the He-burning loop (BHeB), and the
red edge of the HE-burning loop (RHeB). These curves come from the
stellar evolution models of \citet{ber94}. Notice the excellent
agreement between the models and the observations. In particular, the
faint end of the BHeB sequence is now very well detected, and follows
the model prediction with exquisite accuracy. Brighter than $V=22$ the
model prediction and observations diverge slightly. This corresponds
to a stellar mass of about $5_\odot$ and an age of about 100 Myr.

The most obvious difference between the two data sets is the fainter
limits of the Cycle~7 data. The main sequence is fully 2 magnitudes
deeper. Furthermore, the faint limit of the red clump has clearly been
detected, while only the bright end was detected by the Cycle~5 data.
The bright portion of the CMD's are remarkably similar. The MS, BHeB
and RHeB stars all appear in the same positions with similar
densities. In particular, the increased density of BHeB stars near
$V\sim24$ is present in both data sets. The increased quality of the
the Cycle~7 data shows this to be a clear density enhancement. We
shall discuss the implications of this feature in the next section.

\section{Recent Global Star Formation History}

In the following sections we calculate the recent star formation
history (SFH) based on the MS and BHeB populations. To separate these
populations, selection polygons were created. The polygons are based
on the $Z=0.001$ model curves of \citet{ber94} (Fig.\
\ref{figcmd}). The average color index error was used to determine
the width of the polygons as a function of magnitude. Because the
observed BHeB sequence does not follow the model for $V < 22$ a
constant color cutoff of $V-I = -0.1$ was used instead. The selection
polygons are shown in Fig.\ \ref{figcmd}.

Also indicated in Fig.\ \ref{figcmd} is a selection region for the red
giant branch (RGB). We did not include the red clump stars in this
selection in order to obtain the oldest possible star sample. There
are some younger evolutionary phases that may be included, such as the
asymptotic giant branch. However, the RGB region should be dominated
by stars older than 2 Gyr. These stars are not used in this paper for
numerical calculations, so the selection polygon was chosen by
hand. However, care was used to ensure that none of the polygonal
regions overlapped.

\subsection{Main Sequence Stars}
\label{secms}

The luminosity function for the main sequence is shown in Fig.\
\ref{figmslum}. The error bars represent Poisson errors.  The turnoff
age is indicated for several locations.  There are some subtle changes
in slope near $V = 22$, and $V = 24$. Other than this, there are no
discernible features. Fig.\ \ref{figmslum} also shows the MS
luminosity function for the Cycle~5 data as a dotted line. The same
selection region in the CMD was used to isolate the MS in both the
Cycle~7 data and the Cycle~5 data. Other than the photometric depth,
the two histograms are very similar.

Following the prescription of \citet{doh97a} we have calculated the
cumulative mass of stars formed based on the main sequence luminosity
function, and then calculated the star formation rate (SFR) from the
slope of this curve. The result is shown in Fig.\ \ref{figmssfr}.  We
calculated the SFH using two different sets of models to demonstrate
the dependence of the calculation on the model uncertainties. The left
panel shows the calculation using the Padua, $Z=0.001$, stellar
isochrones of \citet{ber94}. The right panel shows the calculation
using the Geneva, $Z=0.001$, stellar evolution models of
\citet{sch92}. In both cases we assumed a power law IMF with a
\citet{sal55} slope ($-1.35$) normalized as described in \citet{doh97a}.
We also normalized all values by the field of view of the WFPC2
instrument. For the adopted distance modulus (25.8) this is an area of
0.90 kpc$^2$. Note that the bins are constant width in magnitude, and
thus vary in width with age.  The error bars represent Poisson errors
based on the total number of stars in each bin and do not include
terms to account for the uncertainties in the stellar evolution
models.

The main feature of the Padua SFH is a strong peak near
25 Myr. The slope change in the luminosity function near $V=22$ is the
rise to this peak, while the slope change near $V=24$ is the sharp
drop-off at 30 Myr. Older than this there is a nearly constant
value, 3-4 times lower than the peak. There is some variation from bin
to bin for these older times, but none of the structure is very
significant within the statistical noise.  For the youngest bins
($<25$ Myr), the sparcity of stars leads to a highly uncertain value,
but the star formation rate generally decreases toward the youngest ages.

The Geneva calculation has the same general form, but the peak is
shifted to near 50 Myr and is not as large in value. However, the
ratio of the peak value (near 25 Myr in the Padua case, and near 50
Myr in the Geneva case) to the off peak values (older than 50 Myr in
the Padua case, and older than 75 Myr in the Geneva case) is the same
for both models. In other words, both models predict a rise in the SFR
in recent times by the same factor. The models disagree in the exact
age of the rise and the actual values of the SFR. The reason for the
disagreement is because the Geneva models predict a slightly lower
mass star, and a slightly older turnoff age, for a given
magnitude. Because of the lower mass, the IMF correction is smaller,
hence the SFR is smaller. Thus, the calculations are demonstrating the
same behavior, there is just disagreement (of order a factor of two)
between the models on the exact normalization values.

Finally, we note that part of the uncertainty in the MS calculation is
endemic to this phase of evolution. Any given location along the MS
can have stars of any age younger than the turnoff age. However, the
situation is even more complicated. The MS stars evolve in luminosity
during the MS phase. Thus, any given magnitude can have stars of
differing mass that may or may not have the same age. Thus, there is
an inherent age and mass resolution limitation on the MS. Because of
this, the MS is not well suited to this type of analysis, where we
calculate the SFH directly from the luminosity function. The MS can be
modeled more accurately with Monte Carlo modeling. The calculation
performed here is primarily for a consistency check on the BHeB
calculation, which we discuss in the next section.

\subsection{Blue He-Burning Stars}

The luminosity function for the BHeB stars is shown in Fig.\
\ref{figbheblum}.  The error bars are calculated as in Fig.\
\ref{figmslum}. The most striking feature is a plateau between
$V=21$ and $V=22.5$. A flattening indicates a reduction in the
SFR. Other than this, the luminosity function is fairly featureless,
except for a few bumps that are not very statistically significant.

For comparison, we have also plotted the BHeB star luminosity function
for the area of the galaxy covered in the Cycle~5 observations.  We
used the same selection region in the CMD for the Cycle~5 data as we
did for the Cycle~7 data. For calculating the SFR from the BHeB stars,
it is not the slope of the luminosity function which is important, but
the actual number count in each bin. The slope indicates the
derivative of the star formation rate with time. The Cycle~5 and
Cycle~7 data do show some differences in slope at various locations,
but, overall, the counts are similar for $V<24$. For fainter
magnitudes, the Cycle~5 data show an excess of stars compared to the
Cycle~7 data. This is because of contamination by MS and red clump
stars, which have scattered into the selection region through
photometric errors (see Fig.\ \ref{figcmd}).  Note that the Cycle~5
luminosity function could be smoother because of the larger
photometric errors at a given luminosity.

From the BHeB luminosity function we have calculated the SFR back to
700 Myr (Fig.\ \ref{figbhebsfr}). The calculation normalizes the star
count in each bin for the IMF, and the length of time stars spend in
the BHeB phase. The details of the calculation can be found in
\citet{doh97a}. We used the $Z=0.001$ stellar evolution models of
\citet{ber94}, and a power law IMF with a \citet{sal55} slope
($-1.35$). Because of model uncertainties in the BHeB phase of very
massive stars ($> 15$ M$_\odot$), we restrict the calculation to lower
masses, and hence ages older than 20 Myr.  We treated the BHeB
identically to the MS stars in that we normalized all values by the
field of view.  Again, the bins are constant width in magnitude, and
thus vary in width with age.

Since the BHeB sequence does change in temperature, with redder colors
at decreasing luminosity, we were concerned that simply binning the
star counts as a function of luminosity might introduce a bias. We
experimented with a method to account for this. Instead of simply
normalizing a star based on its magnitude, we also used the
color index to normalize by the properties of the closest point on the
BHeB model curve in the CMD. We found this made no difference in the
SFR calculation. The reason is that the dynamic range of the BHeB
sequence in color index is much smaller than it is in magnitude, with
the result that accounting for the color index changed the
normalization values very little.

As we saw with the MS (Fig.\ \ref{figmssfr}), the main feature of the
SFH is a peak near 25 Myr. As in the MS SFH, there is an abrupt
drop-off at 30 Myr, and then the SFR decreases with older times, until
leveling out near 100 Myr. Older then 100 Myr the SFR is constant
within the statistical noise, at a level that is roughly 9 times lower
than the peak value. The peak value actually depends somewhat on the
choice of bin size. The actual duration of the star formation event
compared to the bin size is the determining factor. If the bin size is
smaller than the star forming duration, the event is divided among
several bins. If the bin size is larger than the duration, the peak is
smoothed. We have chosen a bin size (0.3 magnitudes) that gives a peak
near maximum value.  Because the calculation was restricted to ages
older than 20 Myr, there is no evidence of the decrease in SFR at the
youngest ages.

Fig.\ \ref{figbhebsfr} also shows the SFH calculation based on the
other half of the galaxy covered in the Cycle~5 data. It is striking
that for ages younger than 300 Myr, the two curves are nearly
identical in form and value. This need not have been the case since
the two data sets cover different portions of the galaxy. This
indicates that, globally, the galaxy seems to have followed the same
history throughout the entire optical body. For ages older than 300
Myr, the Cycle~5 data indicates a higher SFR than the Cycle~7 data. As
discussed above, this excess is likely due to contamination by MS
stars and red clump stars scattering into the selection region.

Finally, we have plotted the SFH based on the Cycle~7 BHeB data
alongside the SFH based on the MS (Fig.\ \ref{figmssfr}) using both
the Padua and Geneva models.  In this case we are hoping for agreement
because it is the SFH of the same region of the galaxy calculated in
two different, and independent, ways.  The Padua models show excellent
agreement in form and value between the MS and BHeB stars.
Specifically, both methods exhibit a strong peak at 25 Myr, with
nearly the same value.  The sharpness of the drop-off at older ages
also appears in both datasets, with the depth of the drop-off
statistically identical.  For older ages ($>30$ Myr), both methods
exhibit a nearly constant, and lower, SFR.  The one notable difference
is that for times older than 100 Myr, the BHeB calculation is
consistently lower than the MS calculation by roughly a factor of two.

The Geneva model's BHeB calculation does not agree quite as well with
the Geneva MS calculation for the youngest ages. As discussed in Sec.\
\ref{secms}, the Geneva MS SFH is shifted to older ages by roughly 25
Myr, and the SFR values are lower, compared to the Padua MS
SFH. However, the MS and BHeB calculations for ages older
than 75 Myr agree better in the Geneva models than they do in the
Padua models. These differences represent systematic model
uncertainities in the SFH calculation, as well as inherent
difficulties in  using the MS for this type of analysis (see Sec.\
\ref{secms}).

Notice that the BHeB calculations of both models (dotted lines in
Fig.\ \ref{figmssfr}) agree with one another extremely well in both
age and value. This indicates the relevant model parameters for the
BHeB evolutionary phase are nearly identical in both models. The BHeB
phase is also not nearly as limited by mass and age resolution as the
MS, which contributes to this agreement. Thus, for this type of
analysis, where the SFR is calculated directly from the luminosity
function, the BHeB stars prove more reliable than the MS.

The excellent agreement between the independent SFR calculations in
the shape and amplitudes of most of the SFR values is a resounding
confirmation for the legitimacy of the using the BHeB stars to
calculate the SFR.  The MS is certainly the best understood of the
phases of stellar evolution, and tying results of the BHeB stars to
those of the MS provides the strongest test of this method.  Due to
brightness limitations, for most galaxies we will not be able to
produce recent star formation histories from MS stars, but the BHeB
stars allow us to study the recent star formation histories of
galaxies beyond the Local Group!

\section{Spatially Resolved Studies of Sextans A}

\subsection{Spatial Distributions of Stellar Populations}

A first cut at a spatially resolved star formation history is to
divide the CMD into broad groups, such as the MS, BHeB, and RGB.  When
discussing the spatial density of stars, it makes sense to include
both the Cycle~5 and Cycle~7 data to obtain a nearly complete view of
the optical galaxy.  In order to orient the reader to these composite
density plots, we include Fig.\ \ref{figupright}.  This is a
reproduction of the HST portion of Fig.\ \ref{figwhole} rotated so the
edges are parallel with the plot axes.  The Cycle~5 image is in the
lower left with the PC toward the upper right.  The Cycle~7 image is
in the upper right rotated $\sim 180$ degrees relative to the Cycle~5
image.

Fig.\ \ref{figden} shows the stellar density for the current positions
of the MS and BHeB stars divided into four different age groups.  The
density maps were made by placing the stars in a spatial grid, and
convolving the grid with a Gaussian kernel. We used $\sigma=80$ pc for
the Gaussian width. For reference, one star per convolution beam is 25
stars kpc$^{-2}$. We divided the luminosity functions into 4 age
groups. One must keep in mind that, for the MS, these groups contain
stars for all ages younger than the labeled age limits. This is not
true of the BHeB stars whose ages are restricted to within the stated
age limits. In both cases, these maps have not been normalized for
IMF, so the numerical values are not directly comparable between
different age groups.

Starting with the upper left hand panel in Fig.\ \ref{figden}, the
youngest MS stars are very concentrated in 3 locations that are about
5 times as dense as the intervening space. These high density
concentrations are the sites of the most recent star formation
activity. They are apparent in the image (Fig.\ \ref{figupright}) as
locations of bright blue stars, and H$\alpha$ emission. These same
concentrations appear in the next oldest category (50-100 Myr) as
well. This is not necessarily because these star forming regions
existed between 50 and 100 Myr ago. Rather, it is because different
age populations overlap along the main sequence such that selecting
stars by the turnoff age includes some stars from all younger
generations as well.

If one accounts for the contamination of younger generations, we see
clear differences between the youngest main sequence density map and
the next older one (50-100 Myr). The highest density regions are now
located in different places.  For example, in the upper left portion
of the figure, the density peak has shifted from $x\sim 450$, $y\sim
1375$ to $x\sim 200$, $y\sim 1125$.  From this we infer that the sites
of star formation have changed over this 50 Myr time scale.

In the oldest two MS star groups, the most obvious change is the lack
of stars in the Cycle~5 portion of the diagram. This simply reflects
the brighter photometric limits of that data set.  In the Cycle~7
portions of the plots, the stars still concentrate, but not as
strongly as in the younger groups. The contrast between the peaks and
intervening space is roughly a factor of 2. Our interpretation is that
the concentrations are stars which are younger than the turnoff limits
we used, and are associated with the youngest groups. The older stars,
present because of the older turnoff limits, are not so concentrated.
The lower contrast can be attributable to either lower star formation
rates or the possibility that these older stars have had time to
disperse from their formation sites.

We now focus on the BHeB stars in the lower panels of Fig.\
\ref{figden} for comparison.  Since the BHeB selection does not mix
ages, they provide a better picture of how the star formation actually
has changed in this galaxy.  In the youngest group, the BHeB stars are
highly concentrated, just as they were for the MS.  As expected, two
of the concentrations coincide with two of the MS concentrations. The
third MS concentration in the youngest group (towards the upper right)
is not present. Since we have restricted the BHeB stars to only those
older than 20 Myr, we infer that this third region is younger than 20
Myr.  Note also that the peak locations in the BHeB stars do not
exactly match up with the peaks in the MS stars. These small
differences are probably due to the 20 Myr restriction on the BHeB
stars. This indicates that the sites of star formation concentrations
can migrate noticeably on time scales as short as 10 Myr.

The differences between the youngest group and the 50-100 Myr group is
quite dramatic. Two strong concentrations, especially the one near
$x=250$, $y=850$, appear without any indication in the younger
group. Furthermore, these prominent concentrations are not present in
the next oldest group (100-150 Myr). This is another indication that
in some cases the timescale for change is much less than 50 Myr.

The oldest two groups still contain concentrations of BHeB stars, but
at much lower contrast to the background, and much lower
signal-to-noise ratio. This may reflect the overall lower SFR, as well
as the diffusion of star forming regions over time.  The appearance of
these concentrations implies that the BHeB stars of this age do retain
at least some spatial structure from their formation, even if they
have begun dispersing.  That is, if there were no concentrations at
these older ages, it would imply that the timescale for the
dispersion of star forming regions is less than roughly 100 Myr.  We
may infer that this timescale is longer than 100 Myr, but the effect
is not strongly constrained by these data due to the low number
counts.  Ideally, one would like to observe a number of star formation
sites of different ages, in order to compare the linear sizes as a
function of time.  Since there are no galaxies that provide high
absolute SFRs over sufficiently long periods of time ($\approx$ 1
Gyr), this type of study will need to be conducted by comparing
similar observations of many different galaxies.

It is interesting to note that there is a lack of BHeB stars in the
upper right portions of the density maps. The extreme upper right
corner is beyond the edge of the optical portion of the galaxy. But
even within this edge there are fewer BHeB stars than on the other
side of the galaxy. It is apparent that over the last 200 Myr this
part of the galaxy has been relatively quiescent. It is only in the
last 20 Myr that we see signs of activity, from the MS.

Finally, we can compare the density of the youngest stars with the
oldest stars in the CMD, namely the RGB. The density of these older
stars is shown in Fig.\ \ref{figrgbden}. They have not been divided
into age groups, as this is a difficult task beyond the scope of this
paper. The most marked difference between these older stars and the
youngest stars is the lack of clumping. Being older, these stars have
had plenty of time to migrate from their birthplace, and distribute
themselves within the galactic potential.  The RGB stars are
centrally concentrated, with a strong bar feature through the center
of the galaxy. We measure the position angle of this feature to be
$49\degr \pm 2\degr$. This matches the major axis of the HI
distribution \citep{sttw88}. Furthermore, \citet{sttw88} perform a
dynamical analysis with a bar potential, and find the projected
morphological axis of the bar to be $52 \pm 3 \degr$, matching well
the observed stellar bar.

It is interesting to note that the distribution of the red clump stars
do not follow this bar structure. The red clump contains stars as
young as 1-2 Gyr. Although we have not yet determined an age
distribution for these stars (to appear in a future paper), we assume
that most of these stars must be younger than the RGB to mask the bar
signature. From this we can infer that it must take at least this long
for stars to relax into the barred potential. It is also interesting
to note that the current star formation does not take place along the
bar. In fact, the highest density HI regions, where stars are
currently forming, are nearly perpendicular to the bar.

\subsection{The Spatially Resolved Star Formation History}

Following the methods of \citet{doh97b} we have created two movies
showing the SFR, as determined by the density of BHeB stars, as a
function of position and time.  The first movie ranges from 20 Myr to
700 Myr in 10 Myr steps. The images have been convolved with an
ellipsoidal Gaussian kernel with $\sigma_{xy} = 80$ pc and $\sigma_t =
30$ Myr.  Still images from the movie are shown in Fig.\
\ref{figmovie}. The movie includes both the Cycle~5 and Cycle~7 data,
and has orientation the same as Fig.\ \ref{figupright}.

By 500 Myr the contrast between the peaks and intervening space is
not as large as in younger ages. There are two reasons for
this. First, for these oldest times, there is some photometric
contamination from other populations, such as the red clump. This is
particularly true of the Cycle~5 (lower left) part of the image, where
the photometric depth does not unambiguously separate the faintest
BHeB stars from other populations. 

A second reason for the smaller contrast may be dynamical. Based on
the density plots above (Fig.\ \ref{figden}) we concluded that the
dynamical timescale for dissolution of a star forming region is at
least 100 Myr, but this number must also have an upper bound, given
the smoothness of the old star distribution. Thus, by 500 Myr, the
star forming regions, while still detectable, are likely to be less
concentrated.  A comparative discussion of the theoretical work
relevant to this question is given in the appendix of \citet{doh97b}.

The linear time resolution is much shorter for the youngest stars than
it is for the older stars. In order to take advantage of this we have
made a second movie ranging from 20 Myr to 100 Myr in 2 Myr time
steps. This movie was also convolved with $\sigma_{xy} = 80$ pc, but
$\sigma_t = 6$ Myr. Still images from this movie are shown in Fig.\
\ref{figmovie100}. 

The first thing to notice about the younger ages is the difference in
transfer function. The SFR ranges from 0 to 15000 M$_\odot$ Myr$^{-1}$
kpc$^{-2}$ as opposed to 0 to 4000 for the older movie. This
emphasizes what is apparent in the global SFH (Fig.\
\ref{figbhebsfr}), which is that the SFR jumps dramatically in the
last 100 Myr. Another feature to notice is that the contrast between
the star forming regions and intervening regions is much higher than
for older times.  The primary reason for this is the higher SFR, which
naturally creates a higher contrast. Furthermore, these regions are
very young and have not yet begun to disperse from their birthplace.

Finally, we wish to point out that there is no {\it concentrated} star
formation in the upper right part of the image until the most recent
times ($< 40$ Myr). It is only in the last 20 Myr that that this part
of the galaxy starts to form stars with comparable vigor. The star
formation associated with the strong H$\alpha$ emission in the upper
right (western) side of the galaxy is clearly very young. This is
somewhat surprising since it is associated with one of the highest
column density gas regions in the galaxy.

By contrast, the prominent group of blue stars in the lower left
(eastern) part of the galaxy has been forming stars for at least 200
Myr. This activity is associated with the highest column density gas
region in the galaxy. The peak location of the activity shifts around,
but always remains near this high density gas cloud. There are also
several lulls in the activity during this 200 Myr time span. So it
appears that this region has been active during several different
episodes over this time period.  This is similar to the pattern of
star formation seen in GR8 by \citet{doh98}.

\subsection{Comparison With Van Dyk et al.\ (1998)}

\citet{van98} performed a comprehensive study of the stellar
population and ISM of Sextans A. Their major conclusion is that an
increase in star formation activity began approximately 50 Myr ago in
the center of the galaxy and progressed outward such that the most
recent activity is taking place on the inner edge of the HI ring. This
model implies that the star formation is taking place in discrete
regions, rather than uniformly distributed throughout the galaxy. They
interpret this to mean that the central activity created a supernova
driven hot bubble which expanded outward.  As the bubble expanded it
compressed the gas and sequentially induced star formation.

A comparison between the location of the youngest MS and BHeB
populations is quite favorable. Comparing their Fig.\ 6 with our Fig.\
\ref{figden} shows strong concentrations of both MS and BHeB stars in
the northern and eastern corners of the optical galaxy. Also apparent
are some BHeB stars in the center of the galaxy. Finally, the youngest
star forming region in the western corner of the galaxy shows a strong
concentration of young MS stars, but almost no BHeB stars. Because of
the lack of BHeB stars in this western region, \citet{van98} also
conclude this is one of the youngest regions of the galaxy, possibly
as young as 15 Myr.

We also find that the central region of the galaxy was more active 50
Myr ago than it is today.  However, the amplitude of the 50 Myr old
central star formation event is not impressive, nor do we see an age
progression from the center outward.  In support of their model,
\citet{van98} highlight the lack of young stars in the center of the
galaxy.  The youngest stars are, instead, found along the inner edge
of the HI ring, and our observations show this as well.  However, the
converse is not true.  We do not find a lack of older stars along the
inner ring of HI.  In particular, the 50 Myr frame of Fig.\
\ref{figmovie100} clearly shows that while the central regions were
forming stars, so were the northern (upper left) and eastern (upper
right) portions of the inner ring edge, at the same rate or higher.
The lack of older stars on the inner edge of the HI ring in the data
of \citet{van98} may be due to the poorer spatial resolution afforded
by their ground based data.  At lower resolution, the brightest stars
of a young population can completely hide an underlying older
(fainter) stellar population.

Thus, our observations do not support the conclusion that the recent
increase in SFR started in the center and progressed outward.  Rather,
it seems to have started in localized regions throughout the galaxy,
but confined to the regions inside the HI ring.  This argues that the
HI ring probably existed prior to 50 Myr ago (which is supported by
our observation of enhanced star formation in the vicinity of the
present day HI concentrations hundreds of Myr ago).  We cannot rule
out the possibility that the star formation in the center of the
galaxy 50 Myr ago cleared away gas which led to enhanced star
formation in the outer parts.  We can only conclude that this was not
the dominant process involved in the recent star formation history of
Sextans A.  Since there is currently a lack of neutral gas in the
central region, it could have been used up by the star formation
activity 50 Myr ago, been heated by this activity, or the ``hole''
could have pre-existed this activity.

\subsection{Neutral Gas Comparison}

We have obtained Sextans~A 21-cm spectral data from the Very Large
Array (VLA) \footnote{The VLA is a telescope of the National Radio
Astronomy Observatory which is operated by Associated Universities,
Ic., under a cooperative agreement with the National Science
Foundation.} archive to create maps of the neutral gas.  The data
consisted of C-array data obtained in May 1992, and D-array data
obtained in July 1992.  Since these observations are used only to show
the relationship between the star forming regions in the current study
and the HI distribution and they are in good agreement with previously
published HI observations \citep{sttw88, van98}, we only briefly
outline their creation below.

All processing steps were performed with standard VLA reduction
software in the NRAO AIPS environment.  The two data sets were first
calibrated, then combined.  The continuum was determined from line
free channels, and subtracted in UV space.  The channel maps were
constructed with the AIPS task IMAGR, using the default (``robust'')
weighting scheme.  This aided in optimizing the trade-off between
signal-to-noise ratio and angular resolution when combining the data
from the two different array configurations.  A process called
conditional transfer was used on the channel maps to remove noise
spikes.  Finally, the flux was converted into gas column density
\citep{gio88}.

The HST images were created from the final, coadded, F555W images, as
described in Sec.\ \ref{obsSec}.  All four chips were combined into a
single image using the IRAF/STSDAS\footnote{IRAF is distributed by the
National Optical Astronomy Observatories, which are operated by the
Association of Universities for Research in Astronomy, Inc., under
cooperative agreement with the National Science Foundation.} routine
wmosaic.  The equatorial coordinates of the HST images were needed to
accurately overlay HI contours.  These were determined using stars
from the USNO-A2 catalog \citep{mon98}.  Both the Cycle~5 and Cycle~7
images each had only a dozen stars available for this transformation.
Because of the low number of stars, only a shift and rotate solution
was determined, rather than a full distortion map. The solution was
found interactively using the IRAF/FINDER package.

The HST images with HI contours are shown in Fig.\ \ref{fighi}.  The
gas has two kidney-shaped high density structures on opposite sides of
the galaxy, connected by a lower density ring.  The optical portion of
the galaxy is coincident with the central HI depression.  The most
recent star formation, including the concentration that is younger
than 20 Myr towards the northwest, lies near the inner edge of the
highest density HI. The fact that the star formation is associated
with the HI is not surprising and has been noted elsewhere
\citep{sttw88, zhsb97, van98}.  What is not so obvious from Fig.\
\ref{fighi}, and may be seen more clearly in Fig.\ 12 of \citet{van98}
is that for the SE HI concentration, the recent star formation
overlaps with highest column density peak, while for the western HI
concentration, the star formation appears to be confined to the inner
edge.  It is also interesting to note that the HI column density is
relatively high across the entire optical disk of the galaxy; that is,
the HI column density never drops below 10$^{20}$ atoms cm$^{-2}$.
Thus, it is probably misleading to refer to a central ``hole'' in the
HI distribution, and more apt to label that feature as a
depression. However, the density of this depression is nearly an order
of magnitude below the nominal critical density for star formation
($10^{21}$ atoms cm$^{-2}$), as discussed in \citet{skh89}.  Note that
the recent star formation in the SW corner of the galaxy is associated
with an HI column density which is comparable to that found in the
central depression.

\section{Discussion}

When discussing the spatial structure of a galaxy, one must ask how
long the features which define a structure will survive, and how their
overall spatial pattern may change with time. A complete discussion of this
for Sextans~A can be found in the appendix of \citet{doh97b}. One of
the most important points to remember for Sextans~A is that the galaxy
is rotating in nearly solid body rotation. Thus, there is no shear to
break apart structures, or to separate gas from stars, although such
separation can occur through other means such as supernova or
wind-driven shocks.

There are three major associations of star formation visible in our
field of view over the past 500 Myr. The oldest is the northern (upper
right) portion of the galaxy. This coincides with an apparent gap
between the two primary HI structures. Stars have been forming in this
region of the galaxy for at least 400 Myr, or longer. The central
location of the star formation does not remain fixed, but migrates
around within this region. The relative lack of gas in this region would
indicate that star formation can not be much longer sustained.

The second oldest region is in the east-southeast portion of the
galaxy. This is associated with the highest column density gas. This
region has been forming stars for at least 200 Myr. Again, the central
location seems to migrate around within this region. 

Finally, the youngest region in the west-northwestern portion of the
galaxy is younger than 20 Myr. There has been no star formation in
this region over the past 700 Myr despite the current presence of the
second highest column density gas structure.

This pattern of star formation leads us to the scenario of an orderly
stochastic process. We envision the star formation as burning through
the gas clouds like a lit fuse. The star formation is induced on the
edge of a cloud. The resulting supernova and wind-driven shocks break
apart the clouds and induce further star formation, progressively
eating their way through the gas structure. 

It remains unclear what induced the rise in SFR over the past
50 Myr. However, given the stochastic nature of the star formation in
these small irregular systems, it may be that this is simply a
positive fluctuation within the expected variation. In fact, it may be
that the galaxy has gone through similar fluctuations in the past 700
Myr, but they are smoothed by limitations in our time resolution.

It is also remarkable that both halves of the galaxy have undergone
the same spatially averaged history, since the local behavior appears
quite different.  Even though the localized star forming regions are
stochastic in nature, it appears that there is some global regulatory
process that controls the average SFR.  Sextans~A is a member of a
group of four galaxies (with Sextans~B, NGC 3109, and the Antlia
dwarf) which are probably bound and lie just outside of the
zero-velocity surface of the Local Group \citep{vdb00}.  With very
similar radial velocities and a separation of only 280 kpc, it is
possible that Sextans A and Sextans B represent a bound pair.  Thus,
it is also possible that the orbital histories of these galaxies
provide a major influence on their star formation histories. Another
possibility is that the barred potential can play some regulatory
role, similar to spiral arms in larger galaxies.

\subsection{Star Formation Models and Future Analysis}

We are reluctant to over-interpret the observations of Sextans~A alone
for two reasons. First, Sextans~A may be in an unusual state compared
to other dI galaxies. For example, Sextans~A has a color index of $B-V
= 0.26$ \citep{hun96} which is much redder than the typical isolated
dI, which has a median value of $B-V = -0.22$ \citep{van00}. In
addition, the large superbubble structure seen in the HII distribution
is unusual for dI galaxies, which typically have centrally
concentrated HII complexes \citep{roy00}.  It is possible that
Sextans~A is going through a rare increase in activity. Such
fluctuations may occur in most, if not all, dI galaxies, but do so
infrequently so that they are not seen in many galaxies. These events
could easily be hidden in the past by the limitations of our time
resolution in determining SFH's.

Second, one would like to observe a large number of star formation
events over time to obtain a coherent picture of the regulatory
processes of star formation. Unfortunately, the size of the galaxy,
and the limited look-back age, allow us only to observe a few such
events, which is not statistically significant.  Since there are no
galaxies that provide high absolute SFRs over sufficiently long
periods of time ($\approx$ 1 Gyr), this type of study will need to be
conducted with similar observations of many different galaxies.  This
will allow a comparative analysis to look for physical characteristics
in the star formation process that are common to all dI galaxies.
Several excellent comparative studies have been done using $H\alpha$
observations \citep{hun98, elm00}. However, these studies provide only
a snapshot of the most recent (a few Myr) activity. This does not
allow one to see how a star forming region changes with time, and link
such changes to physical characteristics of the host galaxy.

Until spatially resolved star formation histories are available for a
large number of dwarf galaxies, it does not make sense to compare
these results with specific models of star formation. Such comparisons
will eventually allow us to characterize such quantities as the size,
duration, efficiency, and regulatory factors of star forming regions
in dwarf galaxies.

\section{Summary}

We have measured stellar photometry from deep Cycle~7, F555W and
F814W, WFPC2 images of the dwarf irregular galaxy Sextans~A. The
photometric measurements are in good agreement with previous, Cycle~5,
WFPC2 observations of a different field in this same galaxy. By
combining the two data sets, we have nearly complete coverage of the
optically visible portion of the galaxy.

The CMD from the new Cycle~7 observations is nearly 2 magnitudes
deeper than the Cycle~5 observations. This allows unambiguous
separation of the faint BHeB stars from contaminants, such as red
clump stars and MS stars. 

We calculated the recent global SFH using both the MS and BHeB
stars. With the photometric limits achieved, we can use the MS to
probe to an age of 200 Myr. This allows us to compare the MS and BHeB
calculations and verify the consistency of the model and method. We
find excellent agreement between the two calculations, both in form
and value. Systematic uncertainties in the distance, IMF, and stellar
evolution models still restrict the precision of the SFR value to a
factor of 2. However, the relative values, and hence the general
pattern of changes in SFR, is much better determined.

The BHeB stars provide a reliable SFR calculation back to 700 Myr.
The global SFH from the Cycle~7 data is remarkably identical to that
of the Cycle~5 data. This implies that both halves of the galaxy have
undergone the same spatially averaged SFH.

We used both the Cycle~5 and Cycle~7 data to examine the spatially
resolved star formation history over the past 700 Myr. The Cycle~5
data are somewhat unreliable for times older than 500 Myr because of
the lower signal-to-noise ratio of these data. We find the stars
frequently formed in localized regions, of order 200 pc in size and
100 Myr in duration. However, these regions are not uniformly
distributed. Rather, the regions are found in 3 major zones of the
galaxy. One of these zones contains star forming regions as old as 400
Myr, and is currently relatively depleted of gas. A second has star
forming regions as old as 200 Myr, and is found on the edge of the
highest column density neutral gas structure. The third has a single
star forming region which is younger than 20 Myr. This youngest region
is on the edge of the second highest column density gas structure.

Our interpretation of this star formation pattern is that the star
forming regions are orderly stochastic events. Star formation is
induced on the edge of gas structures. The subsequent wind and
supernova driven shocks begin to break apart the gas structure, while
inducing further star formation in nearby, but not coincident,
locations. Thus, the star formation eats its way through the gas like
a lit fuse until the gas is either converted into stars, or heated and
dispersed into the galaxy.  A more detailed and quantitative analysis
of the star formation process must await a larger sample of galaxies
with spatially resolved star formation histories.

Finally, we note that it remains a mystery what caused the sudden
increase in SFR over the past 50 Myr. Such fluctuations may be a
natural part of the star formation process, and occur periodically, or
these may have been induced. In fact, short-lived fluctuations may be
hidden in the past by limitations in our time resolution.

\acknowledgements

Support for this work was provided by NASA through grant GO-7496, from
the Space Telescope Science Institute, which is operated by Aura,
Inc., under NASA contract NAS 3-26555. Partial support from NASA
LTSARP grants no. NAGW-3189 and NAG5-9221 and the University of
Minnesota is gratefully acknowledged. This research has made use of
the NASA/IPAC Extragalactic Database (NED) which is operated by the
Jet Propulsion Laboratory, California Institute of Technology, under
contract with NASA.

\begin{deluxetable}{lccc}
\tabletypesize{\scriptsize}
\tablecaption{Sextans~A Cycle~7 WFPC2 Observation Summary}
\tablewidth{0pt}
\tablehead{ & \colhead{F555W} & \colhead{F814W} & 
\colhead{F656N}}
\startdata
Orbits & 8 & 16 & 1\\
Total Exposure Time (sec) & 19200 & 38400 & 2400\\
\enddata
\tablecomments{Listed are the number of orbits obtained in each filter. 
Each orbit consists of two exposures, each 1200 seconds long.}
\label{tabobs}
\end{deluxetable}

\begin{deluxetable}{llcccc}
\tabletypesize{\scriptsize}
\tablecaption{DoPHOT PSF Shape Parameters}
\tablewidth{0pt}
\tablehead{\colhead{Filter} & \colhead{Chip} & \colhead{FWHM} & 
\colhead{$\beta_4$} & \colhead{$\beta_6$} & \colhead{$\beta_8$}}
\startdata
I & 1(PC) & 2.6 & 7.6 &	-8.3 &	3.9 \\
  & 2	  & 1.9 & 3.6 &	-1.1 &	0.4 \\
  & 3	  & 2.1 & 3.6 &	-1.1 &	0.4 \\
  & 4	  & 2.0 & 3.6 &	-1.1 &	0.4 \\
V & 1(PC) & 2.0 & 7.6 &	-8.3 &	3.9 \\
  & 2	  & 1.7 & 3.6 &	-1.1 &	0.4 \\
  & 3	  & 1.9 & 3.6 &	-1.1 &	0.4 \\
  & 4	  & 1.8 & 3.6 &	-1.1 &	0.4 \\
\enddata
\label{tabpsf}
\end{deluxetable}

\begin{deluxetable}{llccccccccc}
\tabletypesize{\scriptsize}
\tablecaption{Aperture Correction}
\tablewidth{0pt}
\tablehead{\colhead{Filter} & \colhead{Chip} & \colhead{$a (10^{-7})$} & 
\colhead{$b (10^{-4})$} & \colhead{$c (10^{-4})$} & \colhead{$d (10^{-7})$} & 
\colhead{$e (10^{-7})$} & \colhead{$f (10^{-7})$} & \colhead{Central Value} & 
\colhead{Stars} & \colhead{RMS}}
\startdata
I & 1(PC) & -12.6 & -0.21 & -4.75 & -1.44 &  3.21 &  2.38 & -0.132 &  323 & 0.149 \\
  & 2     & -7.55 & -1.68 & -1.18 &  0.73 &  2.77 &  0.52 & -0.050 & 1250 & 0.073 \\
  & 3     & -4.39 & -2.50 &  0.55 &  1.00 &  3.62 & -2.14 & -0.039 &  969 & 0.078 \\
  & 4     & -5.71 & -2.11 & -0.14 &  0.50 &  3.60 & -1.62 & -0.050 & 1281 & 0.078 \\
V & 1(PC) & -16.3 & -2.30 & -4.36 &  3.40 & -0.83 &  5.40 & -0.139 &  254 & 0.152 \\
  & 2     & -9.59 & -0.97 & -2.97 & -0.29 &  2.21 &  3.00 & -0.079 &  787 & 0.074 \\
  & 3     & -4.20 & -2.21 &  0.44 &  0.33 &  3.60 & -2.08 & -0.041 &  725 & 0.069 \\
  & 4     & -7.38 & -2.21 & -0.86 &  0.32 &  5.19 & -1.64 & -0.061 & 1170 & 0.080 \\
\enddata
\tablecomments{Correction (mag) $= a + b \times x + c \times y + d \times 
x^2 + e \times x \times y + f \times y^2$}
\label{tabaper}
\end{deluxetable}

\begin{figure}
\caption{The WFPC2 observations of Sextans~A overlaid on a
ground-based image taken by Hunter on the 0.9m at CTIO. North is up
and east to the Left.  The northwestern HST image is from the new
observations described in this paper, while the southeastern image is
described in \citet{doh97a}.  The HST images were created by combining
the F555W ($V$) and F814W ($I$) images. The new observations also
include a F656N (H$\alpha$) image, which is shown in red.}
\label{figwhole}
\end{figure}

\begin{figure}
\caption{The internal DoPHOT error distribution for all 39781 stars
detected in both the F555W ($V$) and F814W ($I$) filters.  The
photometric limit is deeper than the Cycle~5 data by 1.5 mag in $V$
and 1.8 mag in $I$. The distribution is typical of photometric
measurements. The scatter above the main locus is due to stellar
image crowding.}
\label{figerr}
\end{figure}

\begin{figure}
\caption{The completeness fraction based on artificial star tests. The
solid line represents the fraction of stars recovered in $V$, while
the dashed line represents the fraction of stars recovered in $I$. The
error bars indicate Poisson errors based on the number of input
artificial stars. Notice that the fraction of recovered star for
magnitudes between 22 and 26 lies between 80\% and 90\%. This is an
effect of stellar image crowding. The photometric fidelity of the
artificial stars was excellent.}
\label{figcomp}
\end{figure}
    
\begin{figure}
\caption{The color-magnitude diagrams for Sextans~A. The left panel
shows the Cycle~5 data, and the right panel shows the Cycle~7
data. The right axis shows the absolute magnitude for the adopted
distance modulus of $(M-m)=25.8$. The data have been corrected for
Galactic extinction. The most obvious difference between the two data
sets is the increased photometric sensitivity of the Cycle~7 data. The
curves come from the stellar evolution models of \citet{ber94}, and
indicate the main sequence, the blue edge of the He-burning loop, and
the red edge of the He-burning loop. The polygonal regions indicate
the selection regions for the MS, BHeB, and RGB.  Brighter than $V=22$
the color index of the blue and red supergiants differ from the
models. However, for fainter stars the agreement between the model
curves and the data is excellent, particularly for the faint BHeB
sequence.}
\label{figcmd}
\end{figure}

\begin{figure}
\caption{The main sequence luminosity function. The Cycle~7 data are
shown with a solid line and the Cycle~5 data are shown with a dotted
line. The data have been corrected for extinction, and
completeness. The bins are 0.3 magnitudes wide. The errors indicate
both Poisson errors and completeness errors. Near the top of the plot
are indicated the turnoff ages for several values. There are some
subtle changes in slope near $V=22$ and $V=24$, otherwise the histogram
is featureless. The Cycle~5 data does not extend as faint as the
Cycle~7 data, but otherwise the two are very similar.}
\label{figmslum}
\end{figure}

\begin{figure}
\caption{The star formation history based on the main sequence. The
data have been corrected for reddening and incompleteness. The bin
widths are 0.3 magnitudes in $V$, and, hence, vary in width with
age. Error bars include Poisson errors and completeness correction
errors. The left panel shows the calculation using the Padua stellar
isochrones \citep{ber94} to normalize the histogram. The right panel
shows the calculation using the Geneva stellar evolution models
\citep{sch92}.  Also for comparison, the star formation history based
on the BHeB stars has been plotted as a dotted line in both panels
using the respective models. The main feature of the Padua solution is
a strong rise in SFR near 25 Myr, which is shown in both the MS and
BHeB calculations.  For ages greater than 75 Myr, the BHeB calculation
is consistently lower by about a factor of roughly two. The Geneva
models predict a slightly lower mass and turnoff age for a given
magnitude. Thus, we see the peak lowered in value and shifted to
roughly 50 Myr. However, for ages older than 75 Myr, the Geneva model
shows slightly better agreement with the BHeB calculation. Finally,
notice that the BHeB SFH using the Padua models agrees in both age and
value very well with the BHeB SFH using the Geneva models.}
\label{figmssfr}
\end{figure}

\begin{figure}
\caption{The luminosity function of the blue He-burning supergiants
from the Cycle~7 observations. Near the top the age is indicated for
several magnitude values.  The data have been corrected for Galactic
extinction and incompleteness. The bins are 0.3 magnitudes. The error
bars contain both Poisson errors and errors from the incompleteness
correction. For comparison, the Cycle~5 data has been plotted with a
dotted line. There is a plateau between $V=21$ and $V=22.5$ indicating
a reduction in the SFR.}
\label{figbheblum}
\end{figure}

\begin{figure}
\caption{The global star formation history of Sextans~A over the past
700 Myr. The calculation based on the Cycle~7 data is shown with a
solid line. For comparison, the calculation based on the Cycle~5 data
is shown with a dotted line. The calculations used the $Z=0.001$
stellar evolution models of \citet{ber94}, and a power law IMF with a
\citet{sal55} slope ($-1.35$). The bins are 0.3 mag wide in $V$, and
hence vary in width with age. As was true in Fig.\ \ref{figmssfr} the
main feature is an increase in SFR at the youngest times. The Cycle~5
and Cycle~7 calculations are nearly identical, indicating that both
halves of the galaxy have undergone the same average history, despite
the details of the star formation events being different. The apparent
enhancement of SFR in the Cycle~5 data over the Cycle~7 data for
oldest times, is due to contamination of the BHeB population from
other populations, such as the red clump.}
\label{figbhebsfr}
\end{figure}

\begin{figure}
\caption{This figure is identical to the HST insets of Fig.\ \ref{figwhole} 
except they have been rotated counter-clockwise such that the edges 
are nearly parallel with the axes. The Cycle~5 image is in the lower 
left with the PC towards the upper right. The Cycle~7 image is in the 
upper right, rotated nearly 180 degrees compared to the Cycle~5
image. Where the two images overlap, the Cycle~7 image is shown.}
\label{figupright}
\end{figure}

\begin{figure}
\caption{The stellar density maps for the MS and BHeB stars.  The top
row is the MS density and the bottom row is the BHeB density.  The
data have been divided into 4 different age groups: $20-50$ Myr,
50-100 Myr, 100-150 Myr, and 150-200 Myr. The density is shown in both
color (ranging from 0 to 3500 stars kpc$^{-2}$ for the MS and 0 to 100
stars kpc$^{-2}$ for the BHeB) and contours (with levels of 500, 1000,
1500, 2000, 2500, 3000, and 3500 stars kpc$^{-2}$ for the MS, and 25,
50, 75 and 100 stars kpc$^{-2}$ for the BHeB). The maps were convolved
with a Gaussian with $\sigma = 80$ pc. For reference, one star per
convolution beam is 25 stars kpc$^{-2}$. Because of photometric depth
limitations, the Cycle~5 portion of the maps (lower left) do not
extend beyond 100 Myr. Both the MS and BHeB stars are primarily found
in discrete, concentrated regions, indicating both recent and past
star formation sites. In the BHeB maps, notice how the star forming
regions can appear in one frame but disappear from the next. This
indicates the duration of these star forming regions is of order 50
Myr or less. This is not as apparent in the MS maps because selecting
by age only allows us to isolate stars {\em younger} than the turnoff
age. Hence the older MS maps also contain stars whose ages correspond
to the younger maps.}
\label{figden}
\end{figure}

\begin{figure}
\caption{The stellar density map of the RGB stars. The density is
shown in both grey-scale (ranging from 1400 to 3000 stars kpc$^{-2}$)
and contours (with levels 1400, 1800, 2200, 2600, and 3000 stars
kpc$^{-2}$). The map was convolved with a Gaussian kernel with
$\sigma=80$ pc to match the resolution of Fig.\ \ref{figden}. One star
per beam is 25 stars kpc$^{-2}$. The RGB are centrally concentrated in
a bar structure, and smoothly distributed throughout the galaxy. The
orientation of this bar matches the projected morphological bar
determined from HI kinematics in \citet{sttw88}.}
\label{figrgbden}
\end{figure}

\begin{figure}
\caption{The spatially resolved star formation history of
Sextans~A. The nine panels show selected still frames from the full
movie. The movie was convolved with a Gaussian kernel with $\sigma_{xy}
= 80$ pc, and $\sigma_t = 30$ Myr. The intensity transfer function,
and spatial reference is labeled on the central frame. The orientation
of the frames is the same as Fig.\ \ref{figupright}. The SFR ranges
from 0 to 4000 M$_\odot$ Myr$^{-1}$ kpc$^{-2}$. The frames are not
uniformly distributed in time, rather they were chosen to highlight
peaks and valleys in the SFR. Most of the activity is found on the
left side of the maps, and found in two primary zones: upper left
and lower right. The lower right zone is associated with the highest
column density neutral gas.}
\label{figmovie}
\end{figure}

\begin{figure}
\caption{The spatially resolved star formation history of Sextans~A
over the past 100 Myr. The nine panels show still frames from the full
movie. The movie was convolved with a Gaussian kernel with
$\sigma_{xy} = 80$ pc and $\sigma_t = 6$ Myr. The intensity transfer
function, and spatial reference is labeled on the central frame. The
orientation of the frames is the same as Fig.\ \ref{figupright}. The
SFR ranges from 0 to 15000 M$_\odot$ Myr$^{-1}$ kpc$^{-2}$. The SFR
for these young ages is much higher than it is for the older times
(Fig. \ref{figmovie}). There is no concentrated star formation in the
upper right portion of the maps until the last 20 Myr.  This region,
which coincides with the second highest neutral gas column density, is
clearly the youngest.}
\label{figmovie100}
\end{figure}

\begin{figure}
\caption{The neutral gas column density in both contours over a
grey-scale image of the HST data.  The units of the HI are $10^{20}$
cm$^{-2}$. The Cycle~5 field is toward the southeast, and the deeper
Cycle~7 field is toward the northwest. The main features are two
kidney shaped structures of high column density on opposite sides of
the galaxy. These are connected by a lower column density ring, and
surround a central depression. However, note that the central region
still contains a fairly high column of neutral gas, but is only
depressed relative to the surrounding ring. Nearly the entire optical
portion of the galaxy lies within the central HI depression. }
\label{fighi}
\end{figure}

\end{document}